\documentstyle[epsfig,12pt]{article}
\def\bc{\begin{center}}
\def\ec{\end{center}}
\def\be{\begin{equation}}
\def\ee{\end{equation}}
\def\bea{\begin{eqnarray}}
\def\eea{\end{eqnarray}}
\def\nn{\nonumber}

\def\simge{\ \lower-1.2pt\vbox{\hbox{\rlap{$>$}\lower5pt
\vbox{\hbox{$\sim$}}}}\ }
\begin{document}
\pagestyle{empty} 
\vspace{-0.6in}
\begin{flushright}
CERN-TH/97-188 \\
ROME 97/1175 \\
TUM-HEP-279/97 
\end{flushright}
\vskip 0.5cm
\centerline{\large{\bf{CHARMING-PENGUIN ENHANCED $B$ DECAYS}}}
\vskip 1.0cm
\centerline{M.~Ciuchini$^{1,\star}$, R.~Contino$^2$, E.~Franco$^2$,
G. Martinelli$^{2}$, L.~Silvestrini$^3$}
\centerline{\small $^1$  Theory Division, CERN, 1211 Geneva 23, Switzerland.}
\centerline{\small $^2$ Dipartimento di Fisica, Universit\`a ``La Sapienza" and INFN,}
\centerline{\small Sezione di Roma, P.le A. Moro, I-00185 Rome, Italy.}
\centerline{\small $^3$ Physik Department, Technische Universit\"at M\"unchen,}
\centerline{\small D-85748 Garching, Germany.}
\vskip 1.0cm
\abstract{Long-distance contributions of operators of the effective 
non-leptonic weak
Hamiltonian containing charmed currents have been recently studied.
Penguin-like contractions of these operators, denoted
as charming penguins, have been shown to be relevant for several $B$ decays
into two pseudoscalar mesons. In particular,  they
 are expected to give large enhancements to processes which would be otherwise
Cabibbo suppressed.  Their contributions easily lead to values of
$BR(B^+ \to K^0 \pi^+)$ and $BR(B_d \to K^+ \pi^-)$ of about
$1 \times 10^{-5}$, as recently found by the CLEO collaboration. 
In this paper, we show that such large
branching fractions cannot be obtained without charming penguins.
This holds true irrespectively of the model used to compute
the non-leptonic amplitudes and of the relevant parameters of the
CKM mixing matrix.  We use the experimental
measurements of the $B^+ \to K^0 \pi^+ $ and  $B_d \to K^+ \pi^-$ 
 decay rates to constrain the charming-penguin
amplitudes and to predict  $BR$s for a large set of  other two-body decay
channels where their contributions are also important. 
These include several pseudoscalar-vector and
vector-vector channels, such as $B \to \rho K$ and
 $B \to \rho K^*$,  which have not been measured yet.}
\vskip 2cm
\begin{flushleft} 
CERN-TH/97-188 \\
ROME 97/1175 \\
TUM-HEP-297/97 \\
July 1997 \\
Revised Oct. 1997
\end{flushleft}
\vfill
\noindent \underline{\hspace{2in}}\\
$^\star$ On leave of absence from INFN, Sezione Sanit\`a,
 V.le Regina Elena 299, Rome, Italy.
\eject
\pagestyle{empty}\clearpage
\setcounter{page}{1}
\pagestyle{plain}
\newpage 
\pagestyle{plain} \setcounter{page}{1}
\section{Introduction}
\label{sec:introduction} 
In a recent paper \cite{charming1} it has been shown that 
penguin-like contractions of operators containing charmed quarks,
denoted as charming penguins, 
are able to enhance the $B^+ \to K^0 \pi^+ $ and $B_d \to K^+ \pi^-$
decay rates with respect to the values predicted with factorization.
By assuming reasonable values for the charming-penguin contributions,
the corresponding branching ratios are of the order of 
$(1$--$2) \times 10^{-5}$, larger than the expected
$BR(B_d \to \pi^+ \pi^-)$.
This observation is particularly interesting because, in absence of
charming-penguin diagrams, the $B^+ \to K^0 \pi^+$ and
$B_d \to K^+ \pi^-$ rates turn out to be rather small either because
there is a Cabibbo suppression or because the non-Cabibbo suppressed terms
come from penguin operators which have rather small Wilson
coefficients~\footnote{ This is true
unless the corresponding matrix elements are much larger than those predicted
with factorization.}.
The recent CLEO measurements~\cite{CLEO}
\bea
BR(B_d \to K^+ \pi^-) &=& (1.5 ^{+0.5+0.1}_{-0.4-0.1}\pm 0.1)\times
 10^{-5}\nn \\
BR(B^+\to K^0 \pi^+) &=& (2.3  ^{+ 1.1 +0.2} _{-0.9-0.2} \pm 0.2)
\times 10^{-5} \nn\\
BR(B^+ \to \eta^\prime K^+) &=& (7.1^{+2.5}_{-2.1} \pm  0.9  )\times
 10^{-5}
 \label{eq:exp}\\
BR(B_d \to \eta^\prime K^0) &=& (5.3^{+2.8}_{-2.2} \pm  1.2  )\times
 10^{-5} \nn
\eea
allow an estimate of charming-penguin amplitudes
 and call for more quantitative studies.  
\par In this paper, by using the experimental information obtained  
from the measured decay channels, we determine the parameters of
charming penguins and predict a large set of $BR$s 
which have not been measured yet. The main results of our study, which 
will be described in detail below, are summarized in table~\ref{tab:pred}.
In several cases we find that the value of the $BR$
is strongly enhanced by charming penguins. The most interesting channels
are $B_d \to \rho^- K^+$, $B^+ \to \rho^+ K^0$ and $B_d\to\omega K^{(*)0}$,
for which
the contribution of charming penguins is larger than the theoretical
uncertainties and the corresponding $BR$s are close to the 
present experimental upper limits. This means that they will eventually 
be measured,
and compared to our predictions, in the near future.
Within larger theoretical uncertainties, the $BR$s of
other channels, such as $B\to\pi K^*$, $B^+\to\omega K^{(*)+}$ and 
$B\to\phi K^{(*)}$, are also appreciably enhanced by charming penguins
and close to the present limits.
Large enhancements
are also expected for $B \to \eta^{(\prime)} K$ decays. In these cases, however,
our  predictions are rather poor because of the presence
of contributions, related to the anomaly, which are very difficult to 
evaluate.
For $B_d \to \pi^0 \pi^0$, we typically obtain a $BR$ of about
$(5$--$10) \times 10^{-7}$.  Values as large as $(2$--$3) \times
10^{-6}$ remain, however, an open possibility. Finally, we also
predict the $BR$s of  several $B \to D h$ ($h=\pi$, $\rho$, etc.) 
decay channels, for which only upper bounds exist, see table~\ref{tab:xi2}. 
\par The plan of the paper is the following: in section~\ref{sec:plan},
we recall some basic facts about charming penguins and describe the main
features of the present analysis; in section~\ref{sec:uncertainties} we
discuss the theoretical uncertainties stemming from the parameters of the
 CKM mixing  matrix,  
the choice of the Wilson coefficients of the weak Hamiltonian and the 
models used in the calculation of the relevant amplitudes;
in section~\ref{sec:fixing} the procedure used in the 
determination of the important hadronic parameters is illustrated; 
sections~\ref{sec:predictions} and \ref{sec:conclusions} contain a brief
discussion of the results and the conclusions respectively.
\par For the definition of the various parameters used in this study
and the notation,  the reader should refer to \cite{charming1};
the experimental data have been taken from
refs.~\cite{CLEO,PDG,puntini}.

\section{Charming penguins}
\label{sec:plan}
``Charming penguins" denote  penguin-like contractions in matrix elements of
 operators containing charmed quarks, the contribution
of which  would  vanish using   factorization~\cite{charming1}. 
Their effects  are enhanced in decays where
emission diagrams are Cabibbo suppressed with respect to penguin diagrams. 
 Similar effects can also be obtained by 
a breaking of factorization in the matrix elements of the penguin operators,
$Q_3$--$Q_{10}$, or assuming large chromo-magnetic contributions.
Indeed all penguin-like  effects, including the 
chromo-magnetic ones, have the same quantum numbers as charming penguins.  
For this reason they cannot  be disentangled experimentally. Only 
an explicit non-perturbative theoretical calculation of
the operator matrix elements, done consistently in a given renormalization
scheme (and missing to date), can distinguish the different
terms. On the other hand, our parametrization of charming penguins
effectively accounts for the other penguin effects.
{\scriptsize
\begin{table}[t]
\centering
\begin{tabular}{|l||r|r|r|r|r|r|}
\hline
Channel & QCDSR--CV & QCDSR--BV &Lattice--CV&QM--CV& ABLOPR--CV&Experiment\\ 
& $BR\times 10^5$ &$BR\times 10^5$ & $BR\times 10^5$&
 $BR\times 10^5$& $BR\times 10^5$&  $BR\times 10^5$ \\
\hline 
$B_d\to \eta^\prime K^0$    	& 2.45 [0.19]  	& 1.97--3.55 [0.12--0.67] & 2.49 [0.10]
			 	& 2.39 [0.19]   & 4.48 [0.75]  	  	  & $\star$\\
$B_d\to \eta^\prime K^{*0}$ 	& 0.75 [0.01]  	& 0.30--2.41 [0.00--0.10] & 1.41 [0.00] 	
				& 0.47 [0.01]  	& 2.56 [1.06]  		  & $<9.9$\\
$B_d\to \eta K^0$ 		& 0.07 [0.01]   & 0.01--0.26 [0.00--0.02] & 0.07 [0.00] 
				& 0.06 [0.01]  	& 0.10 [0.02]  		  & --\\
$B_d\to \eta K^{*0}$ 		& 0.04 [0.02] 	& 0.01--0.22 [0.01--0.06] & 0.05 [0.01] 
				& 0.03 [0.02] 	& 0.87 [0.83] 		  & $< 3.3$\\
$B_d\to \omega K^0$ 		& 0.73 [0.04] 	& 0.13--1.70 [0.03--0.38] & 0.80 [0.02] 
				& 0.70 [0.06] 	& 0.64 [0.02] 		  & --\\
$B_d\to \omega K^{*0}$ 		& 1.85 [0.11]   & 1.01--3.26 [0.08--0.58] & 2.76 [0.12] 
				& 1.43 [0.08]  	& 6.26 [0.41]  		  & $< 3.8$\\
$B_d\to \phi K^0$ 		& 1.48 [0.11] 	& 0.68--2.29 [0.10--0.75] & 1.77 [0.08] 
				& 0.92 [0.12]	& 2.59 [0.92] 		  & $< 4.2$\\
$B_d\to \phi K^{*0}$ 		& 3.27 [0.09] 	& 1.90--6.73 [0.07--0.62] & 5.08 [0.11] 
				& 1.65 [0.05]	& 10.3 [0.29] 		  & $<2.2$\\
$B_d\to \pi^0 K^0$ 		& 0.65 [0.04]   & 0.49--0.95 [0.01--0.26] & 0.68 [0.03] 
				& 0.64 [0.03] 	& 0.94 [0.13]  		  & $< 4.1$\\
$B_d\to \pi^0 K^{*0}$ 		& 0.40 [0.03]   & 0.29--0.60 [0.01--0.13] & 0.61 [0.02] 
				& 0.26 [0.03] 	& 0.66 [0.09]  		  & $<2.0$\\
$B_d\to \pi^- K^{*+}$ 		& 3.13 [0.17]   & 2.05--5.07 [0.05--1.01] & 3.47 [0.11]
				& 3.16 [0.16]	& 2.84  [0.44] 		  & $< 6.7$ \\
$B_d\to \rho^- K^+$ 		& 0.63 [0.02] 	& 0.21--0.67 [0.00--0.11] & 1.19 [0.02]
				& 0.53 [0.02]   & 1.89 [0.09]  		  & $< 3.3$ \\
$B_d\to \rho^- K^{*+}$ 		& 2.21 [0.12]   & 1.44--3.57 [0.03--0.72] & 4.15 [0.13] 
				& 1.75 [0.09] 	& 7.57 [1.18]  		  & --\\
$B_d\to \rho^0 K^{0}$ 		& 0.44 [0.03]	& 0.07--1.63 [0.00--0.09] & 0.81 [0.01] 
				& 0.35 [0.03]   & 1.33 [0.43] 		  & $< 3.0$\\
$B_d\to \rho^0 K^{*0}$ 		& 1.01 [0.01]   & 0.50--2.45 [0.01--0.13] & 1.86 [0.02] 
				& 0.60 [0.01] 	& 4.39 [0.16] 		  & $ < 46$\\
$B^+\to \eta^\prime K^+$ 	& 2.31 [0.20]	& 2.05--3.14 [0.11--0.92] & 2.36 [0.10]
				& 2.27 [0.20] 	& 3.80 [0.70] 		  & $\star$\\
$B^+\to \eta^\prime K^{*+}$ 	& 4.14 [0.04]	& 1.45--11.22 [0.00--0.57] & 4.82 [0.02] 
				& 4.18 [0.03] 	& 85.0 [1.59] 		  & $<29$\\
$B^+\to \eta K^+$ 		& 0.12 [0.02] 	& 0.02--0.46 [0.00--0.08] & 0.13 [0.01] 
				& 0.11 [0.01] 	& 0.25 [0.05] 		  & --\\ 
$B^+\to \eta K^{*+}$ 		& 0.27 [0.04] 	& 0.04--1.19 [0.01--0.22] & 0.29 [0.03] 
				& 0.25 [0.04]   & 3.13 [1.09] 		  & $<24$\\
$B^+\to \omega K^+$ 		& 0.60 [0.02] 	& 0.06--1.39 [0.00-0.66]  & 0.65 [0.04] 
				& 0.60 [0.03] 	& 0.39 [0.12] 		  & $\star$\\
$B^+\to \omega K^{*+}$ 		& 1.50 [0.08] 	& 0.84--2.48 [0.03--1.12] & 2.24 [0.16] 
				& 1.20 [0.06] 	& 3.64 [0.59]  		  & $<11$\\
$B^+\to \phi K^+$ 		& 1.54 [0.12]	& 0.70--2.38 [0.10--0.78] & 1.84 [0.09] 
				& 0.96 [0.13]	& 2.69 [0.96]  	  	  & $< 0.53$ \\
$B^+\to \phi K^{*+}$ 		& 3.35 [0.09]   & 1.96--6.90 [0.08--0.64] & 5.21 [0.11] 
				& 1.70 [0.06]   & 10.6 [0.30] 		  & $<4.1 $\\
$B^+\to \pi^0 K^+$ 		& 0.94 [0.18]	& 0.67--1.03 [0.09--0.74] & 0.86 [0.10] 
				& 0.95 [0.18]   & 0.94 [0.45] 		  & $< 1.6$\\
$B^+\to \pi^0 K^{*+}$		& 1.70 [0.14]	& 1.15--2.42 [0.06--0.71] & 1.81 [0.09] 
				& 1.71 [0.13] 	& 1.55 [0.35]   	  & $<8.0$\\
$B^+\to \pi^+ K^{*0}$		& 0.92 [0.13]	& 0.73--1.15 [0.08--0.37] & 1.29 [0.07] 
				& 0.62 [0.12] 	& 1.54 [0.40]   	  & $<3.9$\\
$B^+\to \rho^+ K^0$ 		& 2.67 [0.00]	& 0.88--7.28 [0.00--0.36] & 3.05 [0.00] 
				& 2.62 [0.00]   & 27.3 [0.00]   	  & $<6.4$\\
$B^+\to \rho^+ K^{*0}$ 		& 2.21 [0.09]	& 1.42--4.01 [0.06--0.38] & 3.82 [0.09]
			  	& 1.34 [0.07] 	& 9.50 [1.06]  		  & --\\
$B^+\to \rho^0 K^{+}$ 		& 0.39 [0.06]  	& 0.18--0.61 [0.00--0.21] & 0.61 [0.03] 
				& 0.35 [0.06]	& 0.81 [0.53] 		  & $<1.4$\\
$B^+\to \rho^0 K^{*+}$ 		& 1.26 [0.15]  	& 0.81--1.62 [0.06--0.63] & 2.18 [0.14] 
				& 1.00 [0.11]	& 4.31 [1.18] 		  & $< 90$\\
\hline \hline
$B_d\to \pi^+ \pi^-$ 		& 0.99 [0.63]   & 0.39--2.63 [0.15--1.55] & 0.71 [0.39] 
				& 0.95 [0.60]   & 2.37 [1.71] 		  & $< 1.5$\\
$B_d\to \pi^0 \pi^0$ 		& 0.04 		& 0.01--0.12 [0.01--0.11] & 0.004 [0.002] 
				& 0.06 [0.05]   & 0.01 		  	  & $<0.93$\\
$B_d\to \rho^0 \pi^0$ 		& 0.08 [0.11]   & 0.00--0.21 [0.02--0.21] & 0.01 [0.07] 
				& 0.12 [0.13]   & 0.00 [0.03] 		  & $<1.8$ \\
$B_d\to \rho^+ \pi^-$ 		& 2.67 [1.71]   & 1.02--7.51 [0.45--4.15] & 1.93 [1.07] 
				& 2.55 [1.63]   & 6.29 [4.52]  		  & $<8.8$ \\
$B_d\to \pi^+ \rho^-$ 		& 0.59 [0.38]	& 0.22--1.75 [0.11--0.91] & 0.66 [0.37] 
				& 0.46 [0.30] 	& 1.92 [1.38]  		  &  -- \\
$B_d\to \rho^0 \rho^0$ 		& 0.16 [0.18]  	& 0.00--0.58 [0.04--0.46] & 0.02 [0.18] 
				& 0.21 [0.19]   & 0.10 [0.19]		  & $< 28$  \\
$B_d\to \rho^+ \rho^-$ 		& 1.81 [1.16]  	& 0.69--5.10 [0.31--2.82] & 2.19 [1.22] 
				& 1.37 [0.88] 	& 16.6 [11.9]	  	  & $< 220 $ \\
$B_d\to\omega \pi^0$ 		& 0.06 [0.01] 	& 0.03--0.42 [0.00--0.29] & 0.08 [0.005] 
				& 0.06 [0.02] 	& 0.06 [0.00] 		  & -- \\	
$B_d \to \omega \rho^0 $	& 0.10 [0.01]   & 0.05--1.22 [0.01--0.73] & 0.14 [0.01]
				& 0.08 [0.01]	& 0.36 [0.01] 		  & $< 3.4$  \\
$B_d \to \phi \pi^0$		& 0.00 [0.00]   & 0.00--0.01 [0.00--0.01] & 0.00 [0.00]
				& 0.00 [0.00]	& 0.00 [0.00] 		  & $<0.65$  \\
$B_d \to \eta^\prime \pi^0$	& 0.48 [0.36]   & 0.04--2.14 [0.01--2.91] & 0.21 [0.22]
				& 0.44 [0.33]	& 1.05 [0.98] 		  & $<2.2$  \\
$B^+\to \pi^+ \pi^0$ 		& 0.53  	& 0.16--1.18  		  & 0.36 
				& 0.51  	& 1.07 			  & $<2.0$\\
$B^+\to \rho^+ \pi^0$ 		& 1.12  	& 0.30--2.62  		  & 0.80  
				& 1.05  	& 2.63 			  & $< 7.7$\\
$B^+\to \pi^+ \rho^0$ 		& 0.52 		& 0.19--1.08   		  & 0.47 
				& 0.47  	& 1.01 			  & $<5.8$\\
$B^+\to \rho^+ \rho^0$ 		& 0.94  	& 0.23--2.11  		  & 1.08  
				& 0.73  	& 7.44  		  & $< 100$\\
$B^+ \to \omega \pi^+$		& 0.84 [0.50] 	& 0.38--2.35 [0.08--1.64] & 0.94 [0.45]  
				& 0.70 [0.45] 	& 1.69 [0.99] 		  & $\star$ \\
$B^+ \to \omega \rho^+$		& 1.81 [0.94] 	& 0.80--5.57 [0.19--3.40] & 2.44 [1.04]
				& 1.36 [0.72]	& 12.9 [7.23] 		  & -- \\
$B^+ \to \phi \pi^+$		& 0.00 [0.00] 	& 0.00--0.02 [0.00--0.02] & 0.00 [0.00]
				& 0.00 [0.00]	& 0.00 [0.00] 		  & $<0.56$ \\
$B^+\to \pi^+ \eta^\prime$ 	& 1.09 [0.76]	& 0.01--3.33 [0.13--3.36] & 0.87 [0.58]
				& 0.97 [0.65]   & 3.09 [2.53] 		  & $<4.5$\\
$B_d\to \pi^0 J/\Psi$ 		& 2.27  	& 2.06-2.36 		  & 1.82 
				& 2.70   	& 0.41   		  & $< 6$\\
$B_d\to \rho^0 J/\Psi$ 		& 4.44 		& 4.04--4.63  		  & 5.05 
				& 3.61  	& 1.43  		  & $< 25$\\
$B^+\to \rho^+ J/\Psi$ 		& 9.22  	& 8.39--9.62   		  & 10.48 
				& 7.49 		& 2.97 		 	  & $< 77$\\
$B_d\to K^0 \bar K^0$ 		& 0.09 [0.01] 	& 0.05--0.56 [0.00--0.41] & 0.09 [0.005] 
				& 0.08 [0.01] 	& 0.13 [0.03] 		  & $< 1.7$\\
$B_d\to K^+ K^-$ 		& 0.00 [0.00] 	& 0.00--0.50 [0.00--0.50] & 0.00 [0.00] 
				& 0.00 [0.00] 	& 0.00 [0.00] 		  & $< 1.7$\\
$B^+\to K^+ \bar K^0$ 		& 0.09 [0.01]   & 0.05--0.58 [0.00--0.43] & 0.09 [0.005] 
				& 0.08 [0.01]   & 0.13 [0.03] 		  & $< 2.1$\\
\hline
\end{tabular}
\caption{\it{Predictions for yet unmeasured $BR$s. The channels
above the horizontal line are those enhanced by charming penguins.
In square brackets, the $BR$s obtained without charming penguins are shown.
$BR(B^+\to\omega K^+)$, $BR(B^+\to\omega \pi^+)$,
$BR(B^+\to \eta^\prime K^+)$ and $BR(B_d\to \eta^\prime K^0)$, for which
preliminary measurements exist ($(1.5^{+0.7}_{-0.6}\pm
0.3)\times 10^{-5}$, $(1.1^{+0.6}_{-0.5}\pm 0.2)\times 10^{-5}$,
$(7.1^{+2.5}_{-2.1}\pm 0.9)\times 10^{-5}$ and 
$(5.3^{+2.8}_{-2.2}\pm 1.2)\times 10^{-5}$ respectively),
have been given in
this table since they have not been used in the fit of $\eta_L$ and $\delta_L$.}}
\label{tab:pred}
\end{table}
}
\par
In absence of a consistent calculation of the matrix elements, penguin effects
can also be enhanced by choosing a procedure where large ``effective'' Wilson
coefficients of
the penguin operators are obtained and factorization is used to compute the
hadronic amplitudes. Attempts to
reproduce the experimental data along this line have been recently made, see
for example refs.~\cite{bufle,ag} (the results of ref.~\cite{ag}
will be discussed more in detail in section~\ref{sec:fixing}).
In these approaches, the physical mechanism for the enhancement is similar
to the one of ref.~\cite{charming1}, but the procedure 
used in the calculations  is completely  different:  
penguin diagrams involving charmed quarks   are computed perturbatively
on external quark states, and their effects included in the 
effective Wilson coefficients of the penguin operators $Q_3$--$Q_{10}$. 
It can be shown, however, that perturbation theory cannot be used 
to evaluate these effects: the relevant kinematical range of 
 momenta involved in these calculations~\cite{bufle,ag}
corresponds  to a region  where charm-quark threshold effects, 
which are  largely responsible for the enhancement of the 
$B \to K \pi$ rates,  are important
and the perturbative approach ought to fail~\cite{noifut}.
For these reason, following ref.~\cite{charming1}, we parametrize 
charming-penguin effects in terms of the non-perturbative  parameters
$\eta_L$ and $\delta_L$, rather than make any attempt to compute them using
perturbation theory and factorization.
\begin{table}[t]
\centering
\begin{tabular}{|l||r||r|r|r||r|}
\hline
Channel & QCDSR--CV & Lattice--CV & QM--CV & ABLOPR--CV &  Exp. upper bound\\
& $BR\times 10^5$ & $BR\times 10^5$ & $BR\times 10^5$
&$BR\times 10^5$& $BR\times 10^5$ \\
\hline
&&&&&\\
$B_d\to \pi^0 \bar D^0$    	& 17 	& 11 	& 17 	& 4 	& $< 33$\\
$B_d\to \pi^0 \bar D^{*0}$ 	& 28 	& 19 	& 35 	& 6 	& $< 55$\\
$B_d\to \rho^0 \bar D^0$   	& 8 	& 10 	& 10 	& 3 	& $< 55$\\
$B_d\to \rho^0 \bar D^{*0}$	& 32 	& 40 	& 28 	& 19 	& $< 117$\\
$B_d\to \omega \bar D^0$   	& 8 	& 10 	& 9 	& 3 	& $< 57$\\
$B_d\to \omega \bar D^{*0}$	& 32 	& 40 	& 27 	& 19 	& $< 120$\\
$B_d\to \eta^\prime \bar D^0$   & 8  	& 5 	& 8 	& 2 	& $< 33$\\
$B_d\to \eta^\prime \bar D^{*0}$& 12 	& 8 	& 14 	& 48 	& $< 50$\\
&&&&&\\
\hline
\end{tabular}
\caption{\it{Predictions for emission-dominated decay channels
for which only experimental upper bounds exist.}}
\label{tab:xi2}
\end{table}
\par
We now illustrate the main features of our new analysis:
\begin{itemize}
\item[i)] in ref.~\cite{charming1},
all the amplitudes were normalized to the disconnected emission diagrams,
without computing them in any specific model.  This makes the results
model independent but only allows  predictions of 
ratios of partial rates.  In the present work we compute the
disconnected-emission 
diagrams in the factorization approximation, which allows us 
to predict absolute  branching ratios.
 The values of the hadronic parameters $\xi$ and $\delta_\xi$,
relevant for connected-emission diagrams,
are extracted  from a fit to decay channels which are expected to be  dominated by  emission 
diagrams, namely $B \to\pi D^{(*)}$, $B \to\rho D^{(*)}$,
 $B\to K^{(*)} J/\Psi$ and $B\to D D^{(*)}$;
we also allow for some breaking of factorization, by varying 
the disconnected-emission amplitudes  by $15 \%$ with respect to their
factorized value;
 
\item[ii)] we do not fit the Wilson coefficients of the relevant operators
of the weak Hamiltonian,
in the QCD combinations $a_1$ and $a_2$,
see for example ref.~\cite{nestech}. We rather compute them at leading (LO)
and 
next-to-leading  (NLO) order in perturbation theory, at  several values of
the renormalization  scale $\mu$. In this way we  estimate the uncertainty
given by the scale and renormalization prescription dependence;

\item[iii)] the values of the charming-penguin parameters $\eta_L$ and
$\delta_L$ are  constrained  by fitting the first two  decay channels in
eq.~(\ref{eq:exp}). We do not attempt to fit the $B \to \eta^\prime K$
decay rates since  these channels receive  contributions, due to the anomaly,
which are difficult to estimate.  For these channels we will present in the
following a rough estimate of the branching ratio, obtained by including
non-anomalous contributions only.

\item[iv)] using the values of the parameters as determined in i) and iii),
a large set of  $BR$s for  two-body decay channels are predicted, 
including  pseudoscalar-vector and vector-vector final states which were not
considered
in ref.~\cite{charming1}. The results are also compared with the available 
experimental bounds;

\item[v)] a comparative analysis of two-body $BR$s with
 and without charming penguins is performed. 
We show that there is  strong evidence of a
large charming penguin contribution in the data. As the experimental errors
will be reduced, and more channels will be measured,
 this evidence will eventually find more support;

\item[vi)] given its relevance in the extraction of the CP violating angle
$\alpha$ ~\cite{revs}, 
particular attention is devoted to $BR(B_d\to\pi\pi)$ decay.
We will give an average  value of the $BR(B_d\to\pi^0\pi^0)$, corresponding 
to central values of our parameters, and a maximum value, 
obtained by varying  annihilation, GIM-penguin and charming-penguin 
contributions~\cite{charming1} to this channel, 
with the constraint that the predictions of  the measured channels remain 
compatible with the data at the $2$-$\sigma$ level.
\end{itemize}
\par In our analysis, we are mostly concerned with large enhancements induced 
by charming penguins,
rather than with effects of the order of $20$--$30 \%$. 
Moreover our study is not focused in testing the factorization hypothesis. 
For these reasons,
since  we obtain  without any special effort satisfactory results
for those   decay channels which are dominated by
emission diagrams, we have neither tried to  optimize 
the parameters in this sector  nor to estimate  
an error on the theoretical predictions. For these channels,
we will only present results obtained with different models and Wilson
coefficients. The spread of the results is representative of the theoretical 
error. \par  
On the other hand, the  present experimental
information on charming-penguin dominated decays is not sufficient to fix
precisely the relevant non-perturbative parameters.
This is not a problem specific to our approach:  
in all cases where charming penguins are important,
there are many smallish contributions (due to annihilations, GIM-penguins etc.
\cite{charming1}) 
which cannot be precisely estimated by any non-perturbative method. 
 Thus
we will present a band of expected values for those channels where the
enhancement is very large.

\section{Uncertainties of the theoretical predictions}
\label{sec:uncertainties}
In order to assess the relevance of the charming penguins and make 
sensible predictions, a central issue is checking the stability of the results.
We have studied their dependence on several experimental and theoretical
 parameters which enter the calculation of the decay amplitudes. In particular:
\begin{enumerate}
\item most of the theoretical predictions for two-body decays of heavy mesons
 are based on factorized formulae~\cite{dibart}--\cite{alek}.
Although we do not assume factorization, we normalize the amplitudes to the
factorized values of the disconnected-emission diagrams~\cite{charming1}.
 The latter depend on the particular model used to evaluate the relevant form
factors~\cite{formq}--\cite{forml}. We have selected a set of representative
 models which are
consistent with the scaling laws derived from the HQET, namely
lattice QCD~\cite{forml} (LQCD), the quark
model of ref.~\cite{mel} (QM) and the most recent results from
QCD sum rules~\cite{ball-rucl} (QCDSR).
We also consider the model used in ref.~\cite{alek} (ABLOPR), 
because a detailed analysis of the ratio 
$BR(B_d \to K^+ \pi^-)/BR(B_d \to \pi^+ \pi^-)$
was presented there. We analyse the stability of the values
of the  fitted parameters ($\xi$, $\delta_\xi$, $\eta_L$ and $\delta_L$)
 and of the predicted $BR$s for different choices of the model 
used for computing the form factors; 

\item  we vary the value of CKM-parameter $\sigma$ and the CP violating phase
 $\delta$ ($\sigma e^{-i \delta} = \rho - i \eta$ in the standard Wolfenstein
parametrization~\cite{wolf}).  We choose the ranges $\sigma=0.24$--$0.48$ and
$-0.7 \le \cos\delta \le +0.7$, although it can be argued on the basis of
several analyses~\cite{ciuc,ckmfit} that $\cos\delta \simge 0$;

\item we check the stability of our results against
other effects which are suppressed in the factorization hypothesis. 
The contribution of annihilation diagrams and GIM penguins, as defined 
in ref.~\cite{charming1}, has been studied.
\end{enumerate}

Before presenting our results, it is appropriate to  discuss 
the evaluation of the relevant form factors used in the calculation 
of the disconnected emission amplitudes. 
\par Heavy-heavy matrix 
elements are related to the Isgur-Wise (IW) function $\xi(\omega)$. 
HQET allows the 
calculation of $\xi(\omega=1)$ corresponding to $q^2 \sim q^2_{max}$. In most 
of the cases considered here, 
the range of momenta which is used  is instead  around  $q^2\sim 0$. Since
the experimental  measurements support a linear dependence of $\xi$
on $q^2$, the relevant parameter to evaluate the form
factors is the slope of the IW function $\hat \rho^2$ (a
small  extrapolation in $q^2$ is indeed needed 
since data exist down to rather small value of the momentum transfer). 
We find that the $\chi^2$ of our fit to the non-leptonic $BR$s
is a steep function of $\hat\rho^2$. 
The data can be fitted reasonably well only if $\hat\rho^2=0.55$--$0.75$
(with a preferred value of $0.65$), 
which lies within the allowed experimental range~\cite{gibbons}. For  more
detailed and upgraded discussions on factorization and its tests, see 
\cite{noifut,nestech}. 

\par
\begin{figure}

\begin{center}
\vspace{-2.5cm}
\begin{minipage}[b]{.40\linewidth}
 \centering\epsfig{figure=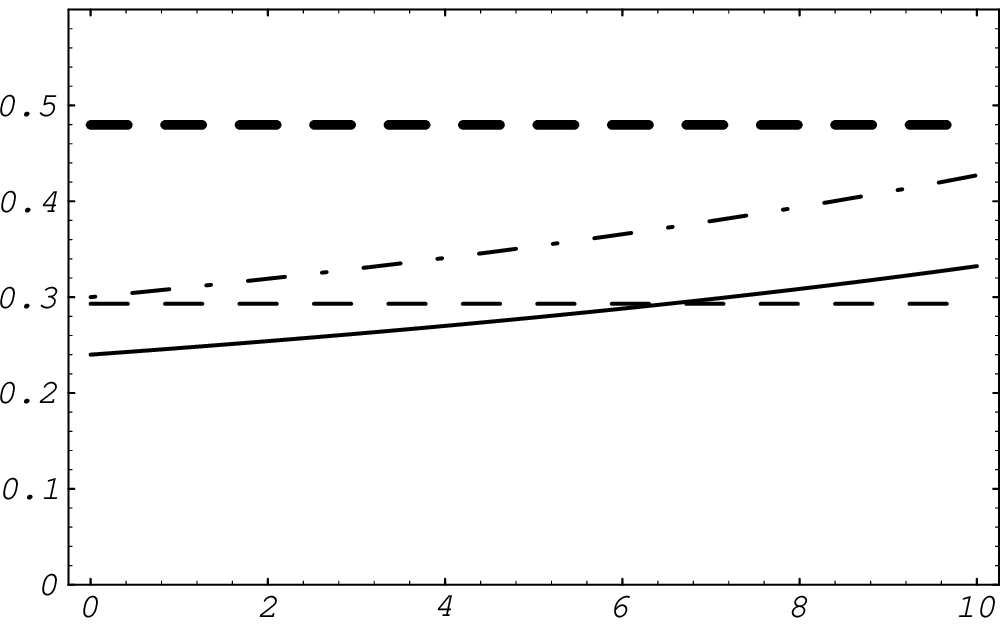,width=\linewidth}
 \vspace{-1.5cm} \\
 $f_{0}$ vs $q^{2}$
\end{minipage}\hfill
\begin{minipage}[b]{.40\linewidth}
 \centering\epsfig{figure=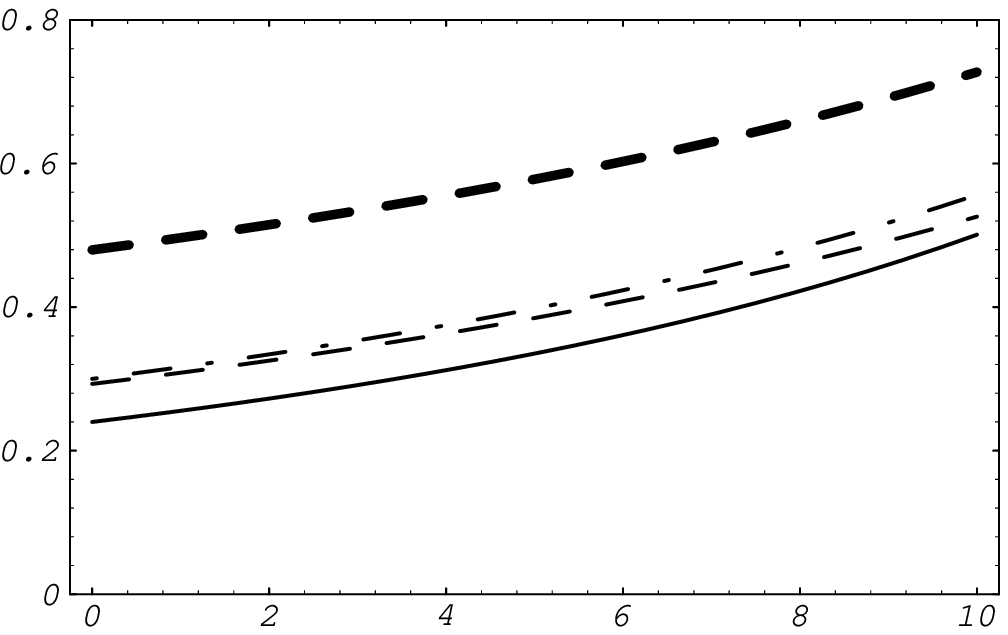,width=\linewidth}
 \vspace{-1.5cm} \\
 $f_{+}$ vs $q^{2}$
\end{minipage}
\vspace{-1cm} \\
\begin{minipage}[b]{.40\linewidth}
 \centering\epsfig{figure=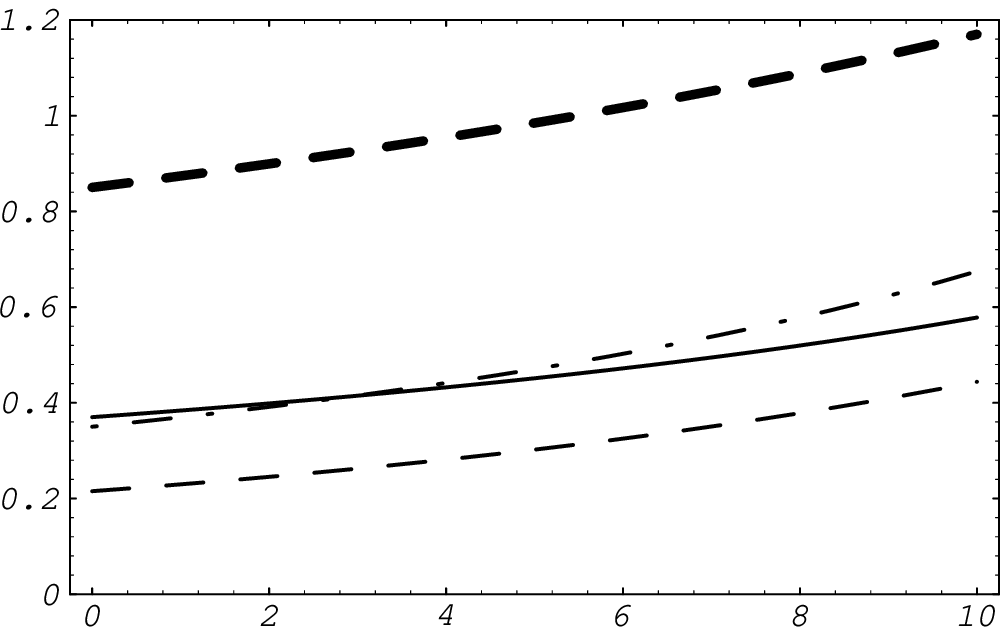,width=\linewidth}
 \vspace{-1.5cm} \\
 $V$ vs $q^{2}$
\end{minipage}\hfill
\begin{minipage}[b]{.40\linewidth}
 \centering\epsfig{figure=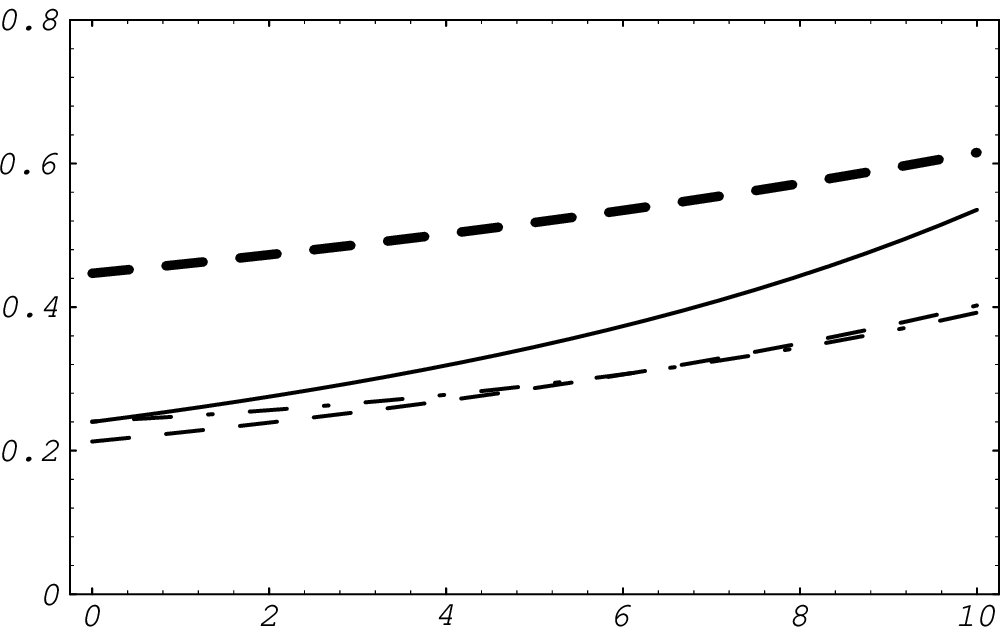,width=\linewidth}
 \vspace{-1.5cm} \\
 $A_{0}$ vs $q^{2}$
\end{minipage}
\vspace{-1cm} \\
\begin{minipage}[b]{.40\linewidth}
 \centering\epsfig{figure=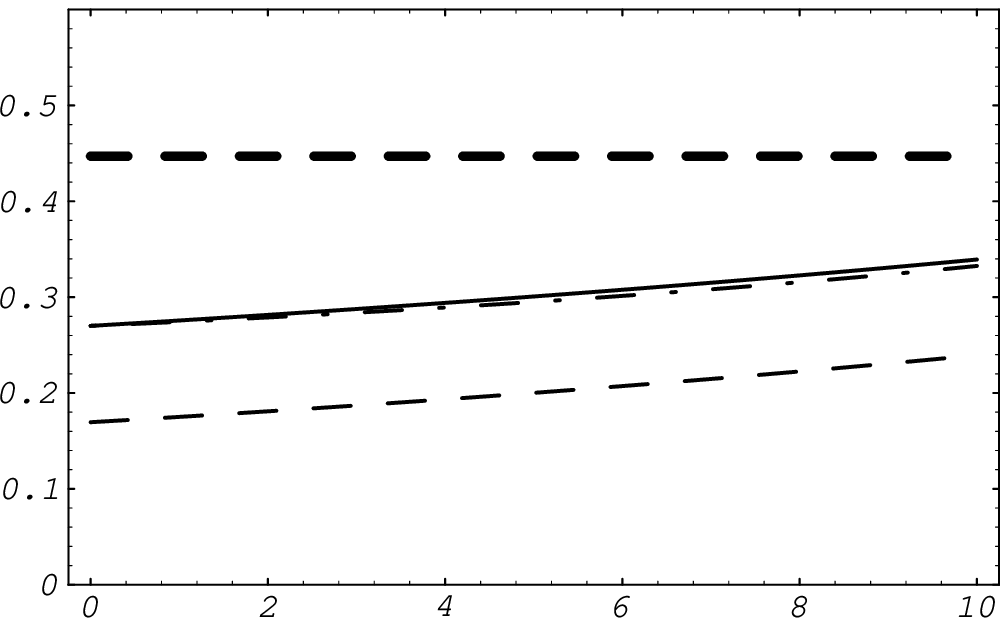,width=\linewidth}
 \vspace{-1.5cm} \\
 $A_{1}$ vs $q^{2}$
\end{minipage} \hfill
\begin{minipage}[b]{.40\linewidth}
 \centering\epsfig{figure=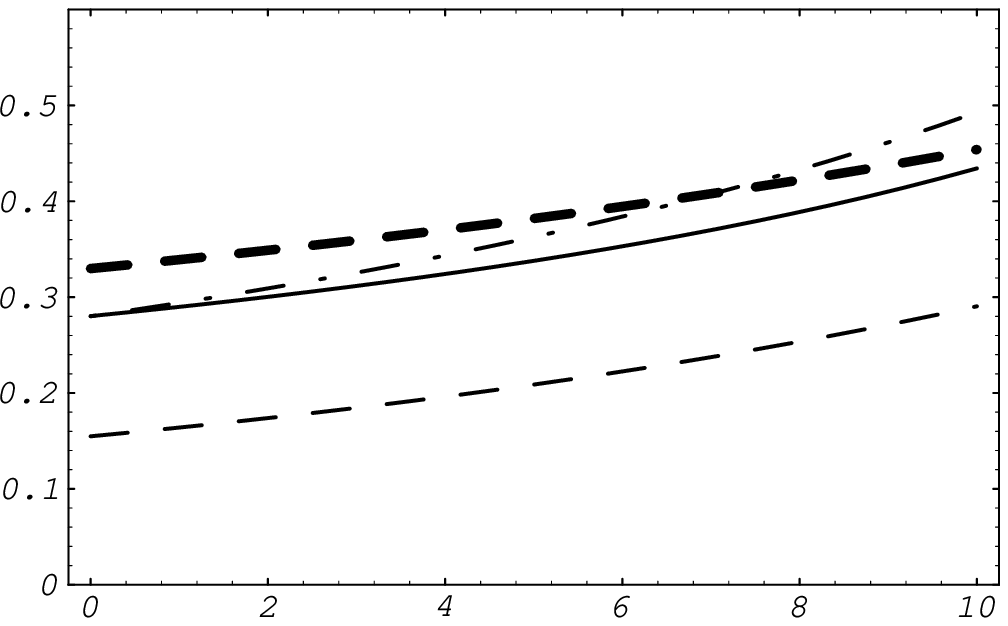,width=\linewidth}
 \vspace{-1.5cm} \\
 $A_{2}$ vs $q^{2}$
\end{minipage} 
\vspace{-1cm} \\
\begin{minipage}[b]{.40\linewidth}
 \centering\epsfig{figure=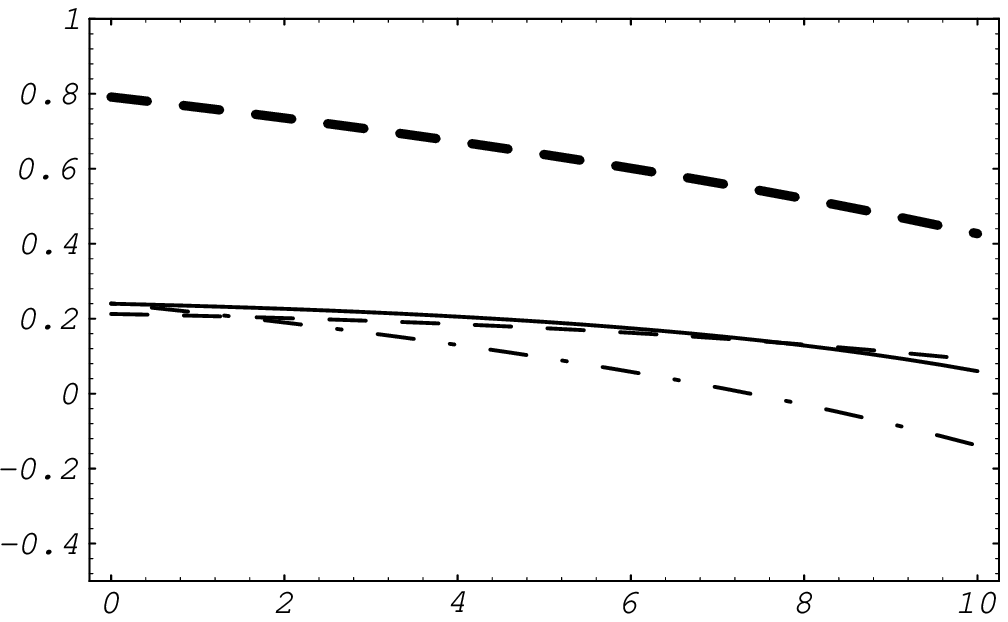,width=\linewidth}
 \vspace{-1.5cm} \\
 $A_{3}$ vs $q^{2}$
\end{minipage}
\end{center}

\caption{\it {Heavy--light form factors vs $q^{2}$ (GeV$^{2}$) from
LQCD (solid), QM (dashed), QCDSR (dot-dashed) and ABLOPR (bold-dashed).}}
\label{fig:ffactors}

\end{figure}

 For the heavy-light form factors,  HQET predicts scaling laws with the 
heavy quark mass only for $q^2 \sim q^2_{max}$~\cite{iwslff}. 
The evaluation of the form factors at the values of $q^2$
needed for the factorized non-leptonic  amplitudes requires further
theoretical inputs, which are provided by a variety of non-perturbative
approaches.  
The state of the art  
 for the set of models used here is  summarized in fig.~\ref{fig:ffactors}.
We find  that the various determinations of the 
heavy-to-light form factors by  the QM of ref.~\cite{mel}, 
QCDSR~\cite{ball-rucl},   and LQCD~\cite{forml}  agree reasonably well 
in a broad range of values of $q^2$. For this reason,
the values of the $BR$s obtained using these model
are very similar, see tables~\ref{tab:pred}, \ref{tab:xi2}
and \ref{tab:xi}. On the contrary,
ABLOPR  systematically  gives 
values of the form factors   $1.5$--$2.5$ times larger  than  the other
models. 
As a consequence most of the $BR$s computed with ABLOPR for $B$ going into
two light mesons are  much larger than in all other cases.
This fact has several consequences that will be discussed in 
sections~\ref{sec:fixing} and \ref{sec:predictions}.

\section{Fixing $\xi$, $\delta_\xi$,
$\eta_L$ and $\delta_L$}
\label{sec:fixing}

We now start the discussion of the results of our analysis. 
We stress again that, given the large 
uncertainties present in the evaluation of the $BR$s, our attitude is to 
focus mainly   on  large effects  induced by charming-penguins. 
Although we do not even try to evaluate the theoretical errors, still we vary 
the input parameters, 
in order to check the stability of our predictions.
{\scriptsize
\begin{table}[t]
\centering
\begin{tabular}{|l||r|r|r|r|r|r|}
\hline
Channel & QCDSR--CV & QCDSR--BV &  Lattice--CV  & QM--CV  & ABLOPR--CV  
& Experiment\\
 & $BR\times 10^{5}$ &  $BR\times 10^{5}$ & $BR\times 10^{5}$ 
& $BR\times 10^{5}$ & $BR\times 10^{5}$ & $BR\times 10^{5}$ \\ 
\hline
&&&&&&\\
$B_d\to \pi^+ D^{-}$ 	  & 301   & 292--344    & 293  & 308  & 318  & $310\pm 44$\\
$B_d\to \pi^+ D^{*-}$ 	  & 323   & 313--369    & 314  & 317  & 341  & $280\pm 41$\\
$B_d\to \rho^+ D^{-}$ 	  & 794   & 770--907    & 771  & 802  & 839  & $840\pm 175$\\
$B_d\to \rho^+ D^{*-}$ 	  & 994   & 964--1136   & 966  & 916  & 1051 & $730\pm 153$\\
$B^+\to \pi^+\bar D^0$ 	  & 508   & 498--527    & 491  & 498  & 437  & $500\pm 54$\\
$B^+\to \pi^+\bar D^{*0}$ & 605   & 595--619    & 595  & 604  & 504  & $520\pm 82$\\
$B^+\to \rho^+\bar D^0$   & 1015  & 988--1103   & 1078 & 1022 & 1015 & $1370\pm 187$\\
$B^+\to \rho^+\bar D^{*0}$& 1396  & 1364---1496 & 1553 & 1269 & 1479 & $1510\pm 301$\\
$B_d\to K^0 J/\Psi$ 	  & 81    & 80--81	& 65   & 96   & 103  & $85\pm 14$\\
$B_d\to K^{*0} J/\Psi$    & 164   & 163--165    & 185  & 120  & 51   & $132\pm 24$\\
$B^+\to K^+ J/\Psi$ 	  &84     & 84    	& 68   & 100  & 107  & $102\pm 11$\\
$B^+\to K^{*+} J/\Psi$    & 171   & 170--172    & 192  & 126  & 54   & $141\pm 33$\\
&&&&&&\\
\hline 
&&&&&&\\
$B_d\to D_s^{+} D^-$	     & 816 [880] 	& 710--1202 [769--940] 	  & 801 [855]  
 			     & 792 [864]  	& 935 [931]       	  & $740\pm 284$\\
$B_d\to D_s^{*+} D^-$ 	     & 824 [887] 	& 723--1253 [844--996]    & 807  [861]
 			     & 802 [871] 	& 842 [938]        	  & $1140\pm 505$\\
$B_d\to D_s^{+} D^{*-}$      & 643 [691] 	& 565--994 [677--796] 	  & 629 [671] 
 			     & 626 [579]	& 734 [731] 	      	  & $940\pm 332$\\
$B_d\to D_s^{*+} D^{*-}$     & 2358 [2535] 	& 2067--3584 [2412--2848] & 2309 [2462]  
			     & 1992 [2165] 	& 2693 [2682] 		  & $2000\pm 729$\\
$B^+\to D_s^{+} \bar D^0$    & 847  [913]  	& 736--1247 [797--975]    & 831 [887]  
			     & 822 [896] 	& 970 [966] 		  & $1360\pm 433$\\
$B^+\to D_s^{*+} \bar D^0$   & 857 [922]   	& 752--1304 [877--1035]   & 840 [895]  
			     & 833 [906]   	& 979 [975] 		  & $940\pm 386$\\
$B^+\to D_s^{+} \bar D^{*0}$ & 669 [718] 	& 588--1034 [704--828] 	  & 655 [697]  
			     & 651 [706]        &763 [760] 		  & $1180\pm 462$\\
$B^+\to D_s^{*+} \bar D^{*0}$&2448 [2632]  	& 2146--3721 [2505--2956] & 2397 [2556]
	              	     & 2067 [2247]      & 2796 [2785] 		  &  $2700\pm 1045$\\
&&&&&&\\
\hline
\multicolumn{7}{|c|}{Results of the fit}\\
\hline
$\xi$ 		& $0.47$  & $(0.40$--$0.57)$ 	& $0.50$  & $0.50$ 	& 0.29 &  \\
$\delta_\xi$ 	& $0.42$  & $(0.21$--$0.68)$ 	& $0.00$  & $0.53$  	& 0.00 &  \\
$\chi^2$/dof 	& 1.00 	   & 0.99--1.63  	& 1.64     & 0.67 	& 1.97  &  \\
\hline
\end{tabular}
\caption{\it{Theoretical predictions for several $BR$s obtained by using
different models for the semileptonic heavy-heavy and heavy-light
form factors. For each model, $\xi$ and $\delta_\xi$ are fitted by minimizing
the $\chi^2$/dof computed using the $BR$s  above the horizontal line only. 
The fitted values of these parameters are also shown, together
with the corresponding  $\chi^2$/dof. For QCDSR-CV, we also give in parentheses
the value of $\xi$ and $\chi^2$ obtained by putting $\delta_\xi=0$. Charming 
penguin contributions, using $\eta_L$ and $\delta_L$ as determined from
 $BR(B^+ \to K^0 \pi^+ )$ and $BR(B_d \to K^+ \pi^-)$, are included.
In square brackets, the $BR$s obtained without charming penguins
are shown.}}
\label{tab:xi}
\end{table}
}
\par
The decay channels which have been used to fix the most
important hadronic parameters  are presented  in tables 
\ref{tab:xi} and \ref{tab:cpmes}. In each table we show a list of
decay channels calculated with and without charming penguins [in square
brackets]~\footnote{ Only in those cases where charming-penguin contractions
contribute.}. 
 The parameters of the charming-penguin contributions,
i.e. $\eta_L$ and $\delta_L$ have been chosen as explained further on in this
section. 
For each decay, we consider  the results obtained using  
different models for the form factors, 
namely the QM  of ref.~\cite{mel}, QCDSR\cite{ball-rucl}, LQCD~\cite{forml} 
and  ABLOPR~\cite{alek}.
The spread of the results obtained in the different cases can be taken as
 an estimate of
the error due to the model dependence.  These results have been obtained
for ``central values'' of the parameters: the Wilson 
coefficients are computed at the LO at a scale $\mu=5$ GeV 
(these coefficients are taken from ref.~\cite{ciuc} 
to which the reader can  refer for details); 
$\cos \delta=0.38$ and $\sigma=0.36$;  
the  disconnected emission diagrams are computed using  factorization; 
GIM-penguin and  annihilation contributions are put to zero. In the following
the results obtained with  this choice of the parameters will be denoted as CV 
(Central Values). 

Table \ref{tab:xi} contains measured decay channels that 
are dominated by emission diagrams (and for which the dominant
contributions are of $O(\lambda^2)$), where $\lambda$ is the Cabibbo
angle in the Wolfenstein parametrization.  They are used for fitting $\xi$ and 
$\delta_\xi$.  Reasonable values of the parameters
are obtained in all cases, with fairly good $\chi^2$ values.
The channels dominating the fit, namely those with the  smaller relative 
experimental error (the $BR$s of which are given 
above the horizontal line in the table),  are 
essentially  independent of  the  charming-penguin
 contributions and of the  CKM parameters $\cos\delta$ and
$\sigma$. For this reason, both the $\chi^2$ and 
the fitted parameters have negligible dependence on these quantities.
The experimental values of the $BR$s, as well as the values of the $\chi^2$ 
and the results of the fit, for each model, are also given.
For the  different  models, different 
determinations of $\xi$ and $\delta_\xi$ are obtained. The larger 
ABLOPR form factors prefer an appreciably smaller value of $\xi$
and have some problem in  reproducing the measured $BR(B\to K^{*} J/\Psi)$.

In table~\ref{tab:xi},
for completeness,   we also
present (for QCDSR only) a band of results obtained in the following way:
we vary  at the LO  the  renormalization scale from $\mu=5$ to $\mu=2$ GeV; 
we  take the Wilson  coefficients computed at the NLO in the NDR scheme 
(but only at $\mu=5$ GeV); we change
the value of the disconnected-emission diagrams by $\pm 15 \%$ with 
respect to their factorized value; we allow 
the value of the GIM-penguin and of the annihilation diagrams 
to vary within the  intervals defined in ref.~\cite{charming1} and take
$-0.7 \le \cos \delta \le +0.7 $ and $0.24 \le \sigma
\le 0.48$.  The spread of values given in the table is representative
of the theoretical uncertainty due to the parameters listed above. 
The band of results   obtained by varying the 
 parameters as explained above is denoted 
as the Band Value (BV).

With the exception of ABLOPR, very similar results are obtained with all the
models considered in our study. For this reason, in table~\ref{tab:cpmes} we
only discuss
the results obtained using QCDSR, as a representative case, and ABLOPR.
In this table, we give a detailed ``map''
of the results obtained  for $BR(B^+ \to K^0 \pi^+ )$
and $BR(B_d \to K^+ \pi^-)$, which have been  used to fit 
the charming-penguin parameters $\eta_L$ and $\delta_L$.
We also present, for the same values of the parameters, $BR(B_d \to
\pi^+ \pi^-)$. The values in square brackets are computed 
without charming
penguins. To illustrate the effects of a large coefficient for the
most important penguin operator $Q_6$, we also give the results  
with the coefficients computed at the NLO in NDR: in this regularization
scheme the relevant penguin coefficients turn out to be larger at NLO than at
LO.  
The values of $\xi$ and $\delta_\xi$ are taken from the fit
to the data of table~\ref{tab:xi}, which only involve heavy-heavy and
heavy-light meson final states. For light-light meson
final states,  it would have been possible, of course,
to choose values of $\xi$ and $\delta_\xi$ different from those fitted 
on the heavy-heavy and heavy-light decay channels.
Different choices of $\xi$ for pseudoscalar-pseudoscalar,
vector-pseudoscalar and vector-vector final states are also possible in
principle. The inflation of parameters would have only complicated the
analysis without changing the basic conclusions.
{\scriptsize 
\begin{table}[t]
\centering
\begin{tabular}{|l||r|r|r|r|r|r|r|}
\hline
 & $\cos\delta$ &$\sigma$ & $BR(B^+ \to K^0 \pi^+ )$ 
& $BR(B_d \to K^+ \pi^-)$ & $BR(B_d \to  \pi^+ \pi^-)$& $\eta_L$
& $\delta_L$  \\
& &  &$BR\times 10^5$   & $BR\times 10^{5}$ &  $BR\times 10^5$ & & \\ 
\hline \hline
Experiment
&  & & $2.3^{+1.1+0.2}_{-0.9-0.2}\pm 0.2 $ & $1.5^{+0.5+0.1}_{-0.4-0.1}\pm0.1 $
 & $< 1.5$ & & \\
 \hline \hline
&&&&&&&\\[-0.1cm]
 &  -0.7 & 0.48 & 1.49 [0.18] & 1.61 [0.48] & 2.09 [0.93] & -0.26 & 1.73 \\
 &  0.38 & 0.48 & 1.62 [0.19] & 1.60 [0.19] & 1.59 [1.10] & -0.29 & 0.38 \\
 &  +0.7 & 0.48 & 1.68 [0.20] & 1.59 [0.11] & 1.49 [1.15] & -0.30 & 0.06 \\
 &  -0.7 & 0.36 & 1.36 [0.19] & 1.62 [0.40] & 1.30 [0.50] & -0.25 & 1.71 \\
QCDSR--CV
 &  0.38 & 0.36 & 1.49 [0.19] & 1.61 [0.19] & 0.99 [0.63] & -0.28 & 0.32 \\
LO $\mu=5$ GeV 
 &  +0.7 & 0.36 & 1.54 [0.20] & 1.61 [0.12] & 0.94 [0.67] & -0.29 & 0.004 \\
 &  -0.7 & 0.24 & 1.23 [0.19] & 1.63 [0.33] & 0.69 [0.21] & -0.24 & 1.62 \\
 &  0.38 & 0.24 & 1.37 [0.19] & 1.62 [0.19] & 0.52 [0.30] & -0.27 & 0.18 \\
 &  +0.7 & 0.24 & 1.42 [0.19] & 1.62 [0.14] & 0.51 [0.32] & -0.27 & -0.10 \\ 
&&&&&&&\\
 \hline
&&&&&&&\\
 &  -0.7 & 0.48 & 1.43 [0.53] & 1.62 [0.97] & 2.24 [0.96] & -0.34 & 1.43 \\
 &  0.38 & 0.48 & 1.75 [0.56] & 1.58 [0.49] & 1.71 [1.26] & -0.28 & 0.25 \\ 
 &  +0.7 & 0.48 & 1.85 [0.57] & 1.57 [0.35] & 1.64 [1.34] & -0.26 & -0.08 \\
 &  -0.7 & 0.36 & 1.30 [0.53] & 1.63 [0.85] & 1.30 [0.51] & -0.32 & 1.25 \\
QCDSR--CV
 &  0.38 & 0.36 & 1.62 [0.56] & 1.60 [0.49] & 1.04 [0.73] & -0.25 & 0.15 \\
NLO $\mu=5$ GeV
 &  +0.7 & 0.36 & 1.71 [0.56] & 1.59 [0.39] & 1.02 [0.80] & -0.24 & -0.16 \\
 &  -0.7 & 0.24 & 1.22 [0.54] & 1.63 [0.75] & 0.47 [0.20] & -0.25 & 0.65 \\
 &  0.38 & 0.24 & 1.50 [0.55] & 1.61 [0.51] & 0.53 [0.35] & -0.22 & -0.4 \\
 &  +0.7 & 0.24 & 1.57 [0.56] & 1.60 [0.44] & 0.54 [0.40] & -0.21 & -0.31 \\
&&&&&&&\\
\hline\hline 
&&&&&&&\\
 &  -0.7 & 0.48 & 2.30 [0.54] & 1.50 [1.27] & 4.67 [2.50] & 1.07 & -0.23\\
 &  0.38 & 0.48 & 2.30 [0.57] & 1.50 [0.52] & 3.55 [2.97] & -0.40 & 0.43 \\
 &  +0.7 & 0.48 & 2.30 [0.58] & 1.50 [0.30] & 3.41 [3.11] & -0.43 & 0.14 \\
 &  -0.7 & 0.36 & 2.03 [0.55] & 1.55 [1.05] & 2.81 [1.36] & 1.01  & -0.23 \\
ABLOPR--CV 
 &  0.38 & 0.36 & 2.19 [0.17] & 1.52 [0.49] & 2.37 [1.71] & -0.33 & 0.82 \\
LO $\mu=5$ GeV
 &  +0.7 & 0.36 & 2.23 [0.57] & 1.51 [0.33] & 2.27 [1.81] & -0.38 & 0.50 \\
 &  -0.7 & 0.24 & 1.78 [0.55] & 1.59 [0.87] & 1.41 [0.56] & 0.93 & -0.25 \\
 &  0.38 & 0.24 & 1.93 [0.56] & 1.57 [0.49] & 1.21 [0.80] & 0.28  & -0.89\\
 &  +0.7 & 0.24 & 1.97 [0.57] & 1.56 [0.38] & 1.17 [0.87] & 0.16 & -1.35 \\ 
&&&&&&&\\
\hline
&&&&&&&\\
 &  -0.7 & 0.48 & 2.29 [1.42] & 1.50 [2.50] & 5.09 [2.51] & 1.43 & -0.22\\
 &  0.38 & 0.48 & 2.30 [1.50] & 1.50 [1.26] & 3.71 [3.29] & -0.24 & 0.64 \\ 
 &  +0.7 & 0.48 & 2.30 [1.52] & 1.50 [0.89] & 3.60 [3.52] & -0.18 & -0.06 \\
 &  -0.7 & 0.36 & 2.01 [1.43] & 1.55 [2.20] & 3.03 [1.34] & 1.34 & -0.23 \\
ABLOPR--CV 
 &  0.38 & 0.36 & 2.30 [1.49] & 1.50 [1.27] & 2.56 [1.92] & -0.32 & 1.16 \\
NLO $\mu=5$ GeV
 &  +0.7 & 0.36 & 2.30 [1.51] & 1.50 [0.99] & 2.33 [2.09] & -0.22 & 0.51 \\
 &  -0.7 & 0.24 & 1.76 [1.44] & 1.59 [1.93] & 1.49 [0.54] & 1.18 & -0.26 \\
 &  0.38 & 0.24 & 2.03 [1.48] & 1.55 [1.31] & 1.29 [0.92] & 0.40 & -0.68 \\
 &  +0.7 & 0.24 & 2.09 [1.49] & 1.54 [1.12] & 1.25 [1.04] & 0.23  & -0.91 \\
&&&&&&&\\
\hline
\end{tabular}
\caption{\it{ $BR$s for several values of the CKM parameters $\cos \delta$
and $\sigma$ and different choices of the Wilson coefficients. The results
are given for QCDSR and ABLOPR only. Results with LQCD and QM are very
similar to those obtained with QCDSR. In square brackets we give the
results obtained without charming penguins.}}
\label{tab:cpmes}
\end{table}
}
\par
When charming penguins are included, QCDSR (as well as  
with LQCD or the QM), give good results for $BR(B^+ \to K^0 \pi^+ )$
and $BR(B_d \to K^+ \pi^-)$ and at the same time respect the experimental
upper bound on  $BR(B_d \to \pi^+ \pi^-)$.  This is true almost irrespectively
of the choice of the Wilson coefficients 
and  of $\cos \delta$~\footnote{Notice that with or without
charming penguins $BR(B^+ \to K^0 \pi^+ )$ is practically independent
of $\cos \delta$.} and of $\sigma$:  only for large values of $\sigma$,
and very small values of $\cos \delta$, there is a problem with
the upper limit on $BR(B_d \to \pi^+ \pi^-)$.  
 There is a general tendency to have  
$BR(B^+ \to K^0 \pi^+ )$ close to $BR(B_d \to K^+ \pi^-)$ 
instead of larger, as found experimentally. 
Given the  experimental errors (and the uncertainties coming
from GIM-penguins and annihilation diagrams) we believe that 
this difference is at present not significant. As far as the last of the
$BR$s in eq.~(\ref{eq:exp}) is concerned, with charming penguins
we typically obtain
$BR(B^+ \to \eta^\prime K^+) \sim (2.5$--$3.0) \times 10^{-5}$, slightly 
less than one-half of the central experimental value (without charming
penguins $BR(B^+ \to \eta^\prime K^+) \sim (0.2$--$1.5) \times 10^{-5}$).
Since the
difference probably comes from anomalous contributions, which are not directly
related to charming penguins and  are very difficult to evaluate,
this discrepancy does not affects our main conclusions. 

With charming penguins, ABLOPR may find even better agreement with the data,
in that it easily reproduces the pattern  $BR(B^+ \to K^0 \pi^+ )
 \sim 1.5 \, BR(B_d \to K^+ \pi^-)$ in  a wide range of $\cos\delta$ and
$\sigma$.
However, it violates   the bound on  $BR(B_d \to \pi^+ \pi^-)$
unless special  values of the parameters are assumed. If we demand good 
agreement for all the three $BR$s, only a low value of $\sigma$ is acceptable.
In this case, with a larger value of the coefficient of $Q_6$, as 
obtained at the NLO, ABLOPR finds agreement with the data even without 
charming penguins.
 
With the exception of ABLOPR, which we discuss separately below,  
we conclude that charming-penguin contributions are essential
 for reproducing the data.
For all the other models that we have studied (QM, LQCD and QCDSR), 
theoretical estimates obtained using factorization
without charming penguins predict too small  $BR$s.
In particular $BR(B^+\to K^0 \pi^+ )$ is always a factor 
$\sim 4$--$10$ smaller than its  measured value.
Although the experimental errors are still rather large,
we believe that  these data already show  
evidence of substantial
charming-penguin  contributions to  $B\to K\pi$ decays.

ABLOPR  deserves a separate discussion. 
By using some specific set of Wilson coefficients (corresponding to a
large value of the coefficient of the  penguin operator $Q_6$), 
this model  may predict correctly the  experimental values for 
$BR(B_d \to K^+ \pi^-)$ and $BR(B^+\to K^0 \pi^+)$ 
 even without charming-penguin contributions.  This is 
 the case, for example, 
if one computes  the coefficients at the NLO, or at
a low value of the renormalization scale $\mu$.  
The reason is that the values of the form factors 
in ABLOPR are quite atypical,  see fig.~\ref{fig:ffactors}. 
The requirement that the experimental values of 
$BR(B_d \to K^+ \pi^-)$ and $BR(B^+\to K^0 \pi^+)$ are reproduced
without  violating other   experimental bounds,
in particular $BR(B_d \to \pi^+ \pi^-)$, demands, however, a
special choice of  the Wilson
coefficients of the operators of the effective Hamiltonian and of the values 
of $\sigma$, which must  be small, or  $\cos \delta$, which must
be negative (see table~\ref{tab:cpmes}). 
Other problems afflict this model, particularly in the vector
sector. 
For example, many experimental upper bounds, see table~\ref{tab:pred},
are violated by using ABLOPR.
This model has also problems  with semileptonic decays
because it gives a very large
semileptonic $BR(B \to \rho)$ (and a relatively large $BR(B \to \pi)$):
one finds $BR(B \to \rho)/BR(B\to \pi) = 2.98$, to be compared to
the experimental number $1.4^{+0.6}_{-0.4} \pm 0.3 \pm 0.4$~\cite{gibbons}
and, as far as $V_{ub}$ is concerned, $V_{ub}/V_{cb}=0.047$ from $B \to \rho$
and $V_{ub}/V_{cb}=0.070 $ from $B \to \pi$.

At present, we cannot completely exclude that, by using ABLOPR, 
it is possible to find a very special set of parameters satisfying the
following  requirement: large
values of $BR(B_d \to K^+ \pi^-)$ and $BR(B^+\to K^0 \pi^+)$, without
invoking charming-penguin effects, and  
respect of the experimental bounds, notably those on $BR(B_d\to \pi^+\pi^-)$
and on $BR$s of channels containing vector mesons in the final states. 
Considering also the problems of ABLOPR with semileptonic decays,
and the fact that it is so different
from all the other models, we find this interpretation of the data rather
remote.
\par
We now discuss the results of ref.~\cite{ag} where the $B\to K\pi$, $B\to
\eta^\prime K$ and $B\to\pi\pi$ branching ratios have also been studied.
The authors of this paper conclude that is possible
to describe the data by using a specific set of effective coefficients, which
include charm-quark penguin effects computed in perturbation theory, and
factorized hadronic amplitudes.

On the theoretical validity of the approach followed in ref.~\cite{ag},
we have already commented in section~\ref{sec:plan}.
Concerning the phenomenological analysis,
our main observation is that, whenever
(any kind of) penguins are important, the values of the $BR$s are strongly
correlated, as shown in table~\ref{tab:cpmes}. As an example, a very small
value of $\xi$, as required to reproduce $BR(B^+\to \eta^\prime K^+)$, leads
to a violation of the experimental bound on $BR(B_d\to \pi^+\pi^-)$.
For this reason, in this table, we have decided to present
the $BR$s for several choices of the main parameters separately.
From the analysis of ref.~\cite{ag}, one may conclude that, for any given
decay channel, there are points in the parameter space where agreement
between theory 
and data can be achieved. However this is no longer the case when
the correlations among different predictions, induced by using the
{\it same} set of parameters for all the light-light channels,
are taken into account. In particular,
on the basis of our experience, we think that it is very difficult to
reproduce the experimental $BR(B^+\to \eta^\prime K^+)$ without violating
the bound on $BR(B_d\to \pi^+\pi^-)$. Clearly,
extra-contributions coming from the anomaly may change this conclusion.

\section{Predictions for yet unmeasured $BR$s}
\label{sec:predictions}
Having determined  the hadronic parameters from the measured $BR$s, 
we use them to predict those  which have not been measured yet. 
These include many channels where the charming-penguin contributions
are Cabibbo-enhanced with respect to emission diagrams, 
together with some  Cabibbo-suppressed channels (of $O(\lambda^3)$),
which are physically interesting, such as $B\to\pi\pi$ 
and $B\to\pi\eta^\prime$. The results are collected in table~\ref{tab:pred}.
As expected,  charming-penguin-dominated channels are strongly
enhanced: typically  by a factor of  $3$--$5$;  in some case,
the enhancement 
can be as large as one order of magnitude. $B \to\rho K^{(*)}$, 
$B\to \omega K^{(*)}$ and $B\to \eta^{(\prime)} K^{(*)}$ are 
 the channels most  sensitive  to charming-penguin effects.
In some case, such as $B\to \eta K$ and $B_d\to \rho^0 K^0$,  
the $BR$s remain small,  at the level of $10^{-6}$, in spite of the 
large enhancement,  and may  be  difficult to measure.

By looking to the column QCDSR-BV, we observe that the typical
uncertainty is about a factor of 2, although, for some channel,
it  can reach even factors of 10 or more.  For this reason, for the Cabibbo-suppressed
channels (corresponding to the $BR$s below the
horizontal line), the errors are so large that predictions with and without
charming penguins largely overlap.
This (h\'elas!) occurs also in some charming-penguin-enhanced 
channels (the $BR$s of which are given above the horizontal line).
All predictions involving $\eta^\prime$ and $\eta$ suffer from the poor
theoretical control that we have on anomalous contributions, which are
never included in our calculations. 

The best candidates to test charming penguins
are $B \to \rho K$ decays (in the different charge combinations) and
$B_d\to\omega K^{(*)0}$.
For these channels, the enhancement is
larger than the estimated uncertainties, and the expected $BR$s are not
far from the present experimental limits, so that one can expect 
that they will soon be measured. Within larger uncertainties,
other  channels  are also appreciably enhanced by charming penguins, 
e.g. $B\to\pi K^*$, $B^+\to \omega K^{(*)+}$ and
$B\to\phi K^{(*)}$.  
With QCDSR, LQCD and the QM the experimental bounds are generally respected;
in many cases, using  ABLOPR,  they are instead  violated, 
often by a large factor, even in absence of charming-penguin contributions.

The Cabibbo-suppressed $B\to\pi\pi$ decay plays a fundamental role in 
the extraction of unitary-triangle angle $\alpha$. 
In ref.~\cite{charming1}, it has been shown that charming-penguin 
contributions in  $B_d\to\pi^+\pi^-$ decays  do not allow the extraction 
of this parameter using only the asymmetry measured in this 
channel. Still, it may be  possible to extract $\alpha$ 
by using the isospin analysis of ref.~\cite{gronau}. This method
requires  $BR(B_d\to\pi^0\pi^0)$ to be large enough.
In ref.~\cite{charming1}, by normalizing the rates to some conventional
value for $BR(B_d \to \pi^+ \pi^-)$, it was  shown that values of
this $BR$ as large as $(2$--$3) \times 10^{-6}$ were possible.  
In the present case, we typically obtain, within
large uncertainties, an average value of $(5$--$10) \times 10^{-7}$. 
We then include  other contributions, 
such as annihilation diagrams and GIM penguins~\cite{charming1}, and  try 
to maximize the $BR(B_d\to \pi^0\pi^0)$ as a 
function of the hadronic parameters, allowing $\chi^2$-values 
twice  larger than those found in the fit. 
In this case,  by stretching all the parameters, we find that  
a maximum value  $BR(B_d\to \pi^0\pi^0)\sim (2$--$3)\times 10^{-6}$
is still allowed. 
Therefore the extraction of $\alpha$ from the 
$B_d\to\pi^+\pi^-$ CP-asymmetry may be a very difficult, hopefully
not impossible,  task.

Finally, in table \ref{tab:xi2}, our predictions of other 
emission-dominated channels, for which only upper bound have been measured,
are also presented. 
We see that all the models satisfy the bounds without any problem.

\section{Conclusions}
\label{sec:conclusions}
In this paper, by using the present experimental measurements and
constraints, we have determined some non-perturbative hadronic parameters
necessary to predict the $BR$s for two-body $B$-meson decays. 
Using this information, we have predicted a large set of yet unmeasured
$BR$s and discussed the corresponding uncertainties. The latter have been
estimated by using different models, different Wilson coefficients
and by varying the CKM parameters within the ranges allowed by experiments.

We found that, in several
cases, notably $B\to\rho K$, charming-penguin effects are larger than
the theoretical error and the expected value of the corresponding $BR$s
are close to the experimental upper bound. Therefore these channels are
the ideal playground to test non-factorizable charming-penguin effects.

\section*{Acknowledgements}
We thank A.~Ali, P.~Ball, V.~Braun, J.~Charles, 
J.~Flynn, C.~Greub, M.~Neubert, O.~P\`ene and A.~Soni 
for useful discussions on the subject of this paper.
L.S. acknowledges the support of ``Fondazione A. Della Riccia''.
We acknowledge the partial support by the EC contract CHRX-CT93-0132
and by M.U.R.S.T.

\end{document}